\documentclass[conference,letterpaper]{IEEEtran}

\usepackage{amsmath,amsfonts,amssymb,amsthm}
\usepackage{psfrag,paralist}
\usepackage{bbm}
\usepackage{dsfont}

\usepackage{graphicx}
\usepackage{epstopdf}
\usepackage{rotating}

\usepackage{color}

\usepackage{tikz}
\usepackage{pgfplots}

\usetikzlibrary{plotmarks}
\usetikzlibrary{calc}
\usetikzlibrary{shapes,arrows}
\usetikzlibrary{decorations.markings}
\usetikzlibrary{positioning}

\pgfplotsset{compat=1.10}

\usepackage{url}
\usepackage{balance}
\usepackage[utf8]{inputenc}

\DeclareFontFamily{U}{mathx}{\hyphenchar\font45}
\DeclareFontShape{U}{mathx}{m}{n}{<-> mathx10}{}
\DeclareSymbolFont{mathx}{U}{mathx}{m}{n}

\usepackage[hidelinks,bookmarks=false]{hyperref}

\begin{document}
\title{6G Sensing Security: Distributed Game-Theoretic RL for Urban Beamforming and Attacker Detection}
\IEEEoverridecommandlockouts

\author{%
  \IEEEauthorblockN{Parmida Geranmayeh\IEEEauthorrefmark{1} and Onur G\"unl\"u\IEEEauthorrefmark{1}\IEEEauthorrefmark{2}}
    \IEEEauthorblockA{%
    \IEEEauthorrefmark{1}Lehrstuhl f\"ur Nachrichtentechnik, Technische Universit\"at Dortmund, Germany\\
    \{parmida.geranmayeh, onur.guenlue\}@tu-dortmund.de\\
    \IEEEauthorrefmark{2}Information Theory and Security Laboratory (ITSL), Link\"oping University, Sweden
  }
}

\maketitle

\begin{abstract}
	In next-generation networks, communication systems will no longer be limited to data transmission and will be expected to acquire awareness of the surrounding environment. This leads to the concept of integrated sensing and communication (ISAC), where the same wireless infrastructure is used for both communication and environmental sensing. Thus, ISAC enables the system to transmit information efficiently and observe and interpret channel variations and user behavior. Motivated by this capability, this work focuses on detecting an active attacker in an urban environment scenario, where the attacker intentionally manipulates beamforming directions to increase interference and mislead the transmitter into allocating the main lobe of beam toward itself instead of legitimate users. We apply game-theoretic approaches to model the interaction between legitimate users and the attacker, and integrate the resulting utility-based formulation into a reinforcement learning (RL) framework. Simulation results demonstrate that the proposed method effectively addresses security challenges in dynamic 6G ISAC systems.
\end{abstract}


\section{Introduction}
Integrated sensing and communication (ISAC) is seen as a promising technology for future wireless networks \cite{mamaghani2026securing,mukherjee2014principles} investigated from different perspectives, such as waveform design, resource allocation, beamforming, and system performance analysis \cite{OnurRoleofISAC,zou2024energy,he2023full}. Combining sensing and communication functions reduces spectrum congestion between radar and communication systems, while improving spectrum efficiency and decreasing hardware cost \cite{zhang2025research}. Recent research has further highlighted the importance of securing ISAC systems to increase trust in the ISAC-based system outcomes \cite{OnurSecureISACJournal,Onur2026secureISACTutorial}. However, interference management remains a critical challenge. Beamforming is an effective approach for mitigating interference and improving system performance \cite{maheshwari2019flexible} and enhances signal quality, increases coverage, and mitigates co-channel interference. These improvements result in enhanced data rates, reduced latency, and overall increased network capacity. \cite{geranmayeh2024comparison}. However, ISAC introduces additional security challenges \cite{mukherjee2014principles}. Consequently, secure designs, particularly in the area of physical-layer security, have become important. Recent work aims to enhance communication security and sensing performance
\cite{mukherjee2014principles,Onur2026ISACfeedback}. 

Network security research based on incomplete-information games has also achieved considerable progress. For instance, in \cite{lei2018incomplete} an incomplete-information Markov game model incorporating moving attack and moving detection surfaces is proposed. This framework jointly considers defense effectiveness and operational cost, while selecting optimal defense strategies under realistic network conditions. In \cite{jiang2021markov}, a multi-stage Markov signaling game approach for moving target defense is considered, where optimal defense strategies are determined despite uncertain prior information. Moreover, \cite{teofilo2012adapting} highlights the importance of adaptive decision-making strategies in incomplete-information games. Additionally, learning-based frameworks have also been extended to ISAC systems to jointly address security, sensing, and communication performance under dynamic and uncertain environments. Reinforcement learning (RL) provides an effective framework in which agents continuously refine their strategies through interactions with the environment. Thus, recent studies combine incomplete-information games with RL techniques. For instance, in \cite{ren2023cooperative} a collaborative decision-making framework is proposed that combines dynamic incomplete-information games with deep reinforcement learning (DRL) to obtain Bayesian Nash equilibrium strategies for unmanned aerial systems. Furthermore, \cite{yao2024bayesian} develops a Bayesian Q-learning and random game-based framework for cyberattack defense in industrial cyber-physical systems, even when attacker information is unavailable. Similarly, in \cite{mamaghani2026securing}, a secure ISAC system is developed, using a Stackelberg game-theoretic framework combined with DRL to jointly enhance communication security, radar sensing performance, and power efficiency.

In this work, we propose a machine learning-based game-theoretic framework to secure a sensing system against active attackers. A 3GPP-based ray-tracing urban scenario (in Dortmund downtown) is considered to model signal propagation. Next, Bayesian inference is incorporated into the utility functions of Nash and Stackelberg games to model uncertainty regarding attacker presence, illustrating that Bayesian inference yields more accurate attacker detection compared to the Nash approach. A Q-learning-based method is then applied to identify the attacker and mitigate its impact by removing it from the received beamforming process.

\section{System Model and Simulation Setup}

The simulation environment is based on the 3GPP TR 38.901 for the Urban Micro (UMi) scenario in 5G networks \cite{ETSI_TR_138901_v1610}. There are 2 base stations as transmitters (TXs) and 4 receivers (RXs). The carrier frequency is 28 GHz, operating in the FR2 mmWave band, with a bandwidth of 400 MHz. Suppose a malicious RX can provide falsified beam-quality or channel-state feedback. Moreover, the network geometry includes base station heights between 8-23 m and user heights between 1.2-2.2 m. The minimum user-to-base-station distance is 10 m, and the inter-site distance (ISD) is 200 m. Links operate under both Line-of-Sight (LOS) and Non-LOS (NLOS) conditions, where the link state is determined according to the 3GPP LOS probability model \cite{ETSI_TR_138901_v1610}
\begin{equation}
\Pr\nolimits_{\text{LOS}}(d) = \min\left\{\frac{18}{d},1 \right\}\cdot \left( 1 - e^{-\frac{d}{36}} \right) + e^{-\frac{d}{36}},
\end{equation}
where $d$ is the distance between a TX and RX. Conditioned on the link state, the channel is modeled as either LOS or NLOS. The LOS channel is represented as
\begin{equation}
\mathbf{H}_{\mathrm{LOS}}=\sqrt{N_r N_t}\,\alpha\,\mathbf{a}_r(\theta_r)\,\mathbf{a}_t^{H}(\theta_t),
\end{equation}
where $N_t$ and $N_r$ are the numbers of transmit and receive antennas, respectively, $\alpha$ is the complex path gain accounting for path loss and shadowing effects, and $\mathbf{a}_t(\theta_t)$ and $\mathbf{a}_r(\theta_r)$ denote the transmit and receive array response vectors associated with the angles of departure and arrival, respectively. The NLOS channel is modeled as
\begin{equation}
\mathbf{H}_{\mathrm{NLOS}}=\sqrt{\frac{N_r N_t}{L}}\sum_{l=1}^{L}\alpha_l\, \mathbf{a}_r(\theta_{r,l})\mathbf{a}_t^{H}(\theta_{t,l}),
\end{equation}
where $L$ denotes the number of propagation paths, $\alpha_l$ is the complex gain of the $l$-th path, and $\theta_{t,l}$ and $\theta_{r,l}$ represent the corresponding angles of departure and arrival, respectively. We use the following path-loss models \cite{ETSI_TR_138901_v1610} 
\begin{align}
&\text{PL}_{\text{LOS}} = 32.4 + 21\log_{10}(d) + 20\log_{10}\left( f_{c} \right),\\
&\text{PL}_{\text{NLOS}} = 32.4 + 31.9\log_{10}(d) + 20\log_{10}\left( f_{c} \right),
\end{align}
where $f_c$ is the carrier frequency in GHz. Note that under the NLOS condition, path losses are larger due to urban obstacles. To model the radio environment more accurately, the shadow fading effect is also considered, representing the slow changes in received power caused by large environmental obstacles such as buildings. We model shadowing as a log-normal random variable and add it to the path loss models with standard deviation $7$ dB. Similarly, small-scale fading is modeled using the 3GPP NR Clustered Delay Line (CDL) model  \cite{ETSI_TR_138901_v1610}, which accurately simulates multipath effects, including cluster delays, fast fading, and variations in signal phase and amplitude. This model is widely regarded as a reference for 5G NR channels. Uniform Rectangular Array (URA) antennas are employed to enable beamforming and massive multiple-input multiple-output (MIMO) systems. A $4\times8$ array is considered on the TX side and a $2\times2$ array is considered on the RX side. The distance between the antenna elements is half the wavelength, and steering vectors are used to guide the beam at different angles. In this method, for each specific angle in the angular scan range, beamforming weights are calculated as a phase vector so that the signals transmitted from different antenna elements are in phase in the desired direction and attenuated in other directions. The beamforming weight vector in this model is defined as $w = a(f,\theta)$ \cite{mathworks2026phased}.

\begin{figure}[t]
	\centering
	\includegraphics[width=0.85\columnwidth, keepaspectratio]{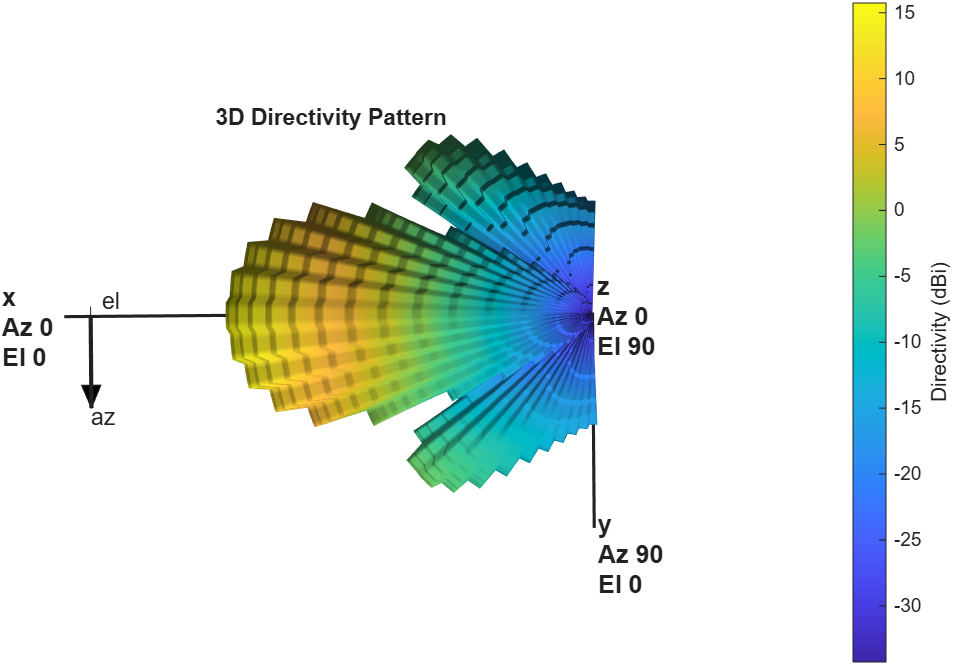}
	\vspace{-0.5cm}
	\caption{The digitized antenna radiation pattern.}
	\label{fig:digitized_antenna_radiation_pattern}
	\vspace{-0.5cm}
\end{figure}

After calculating the weight vector, the array radiation pattern for each angle is calculated by applying beamforming weights to the antenna array and the corresponding beam is generated. Next, to model the communication link, the beamforming gain in the channel is calculated. The effective channel gain is obtained by considering beamforming at the TX and RX sides, defined as in \cite{mathworks2026phased} $G = \left| w_{r}^{H}\mathbf{H}w_{t} \right|^{2}$, where $\mathbf{H}$ denotes the channel matrix obtained from either the selected LOS/NLOS model or, in the ray-tracing simulations, from the resolved Shooting and Bouncing Rays (SBR) propagation paths, and $w_t$ is the beamforming vector at the TX side and $w_r$ at the RX side. Fig. \ref{fig:digitized_antenna_radiation_pattern} illustrates the beam pattern of the RX.

Moreover, thermal noise power is also calculated as \cite{itur2009m2135}
\begin{equation}
P_{\text{noise}}(W) = 10^{\frac{-174 + 10 \log_{10}(B) + NF - 30}{10}},
\end{equation}
where $NF$ is the noise figure of the receiver (in dB) and $B$ is the system bandwidth (in Hz). Finally, the signal-to-noise and interference ratio (SINR) for each user is defined as
$SINR = P_{\text{signal}}/(P_{\text{interference}} + P_{\text{noise}})$, where $P_{\text{signal}}$ and $P_{\text{interference}}$ are the received signal power and aggregate interference power, respectively. The channel capacity of each user is then $ C = B\log_{2}(1 + SINR)$. The effects of path loss, shadow fading, small-scale fading, beamforming, and the CDL channel model are considered in the 5G NR mmWave system, while the urban evaluation additionally uses SBR-based ray tracing. The maximum total channel capacity in the best case in this condition is $264.92$ Mbps.

\begin{figure}[t]
	\centering
	\includegraphics[width=0.8\columnwidth,keepaspectratio]{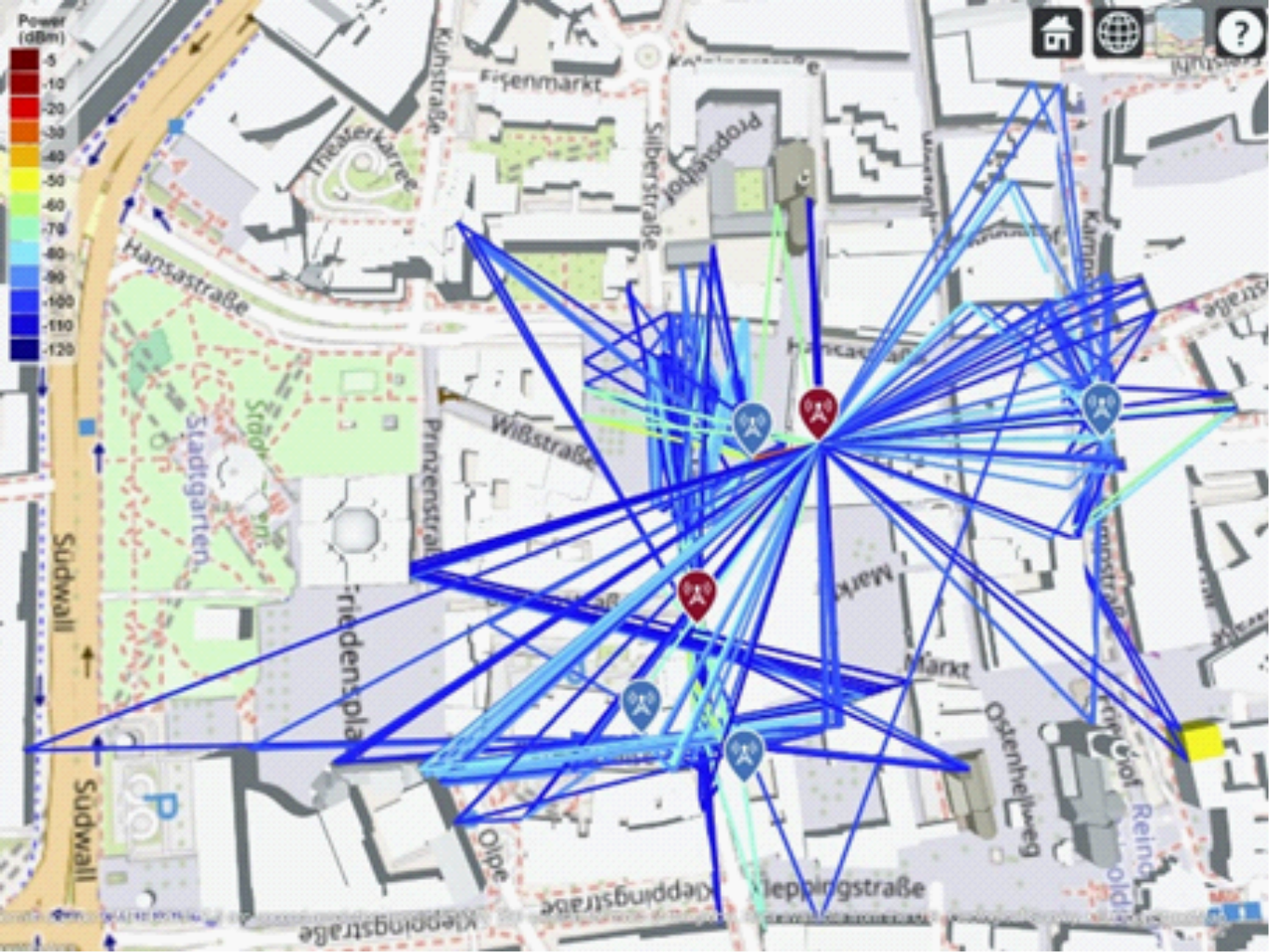}
	\vspace{-0.3cm}
	\caption{Antenna placements in Dortmund downtown.}
	\label{fig:antenna_placement}
	\vspace{-0.1cm}
\end{figure}

The simulations using beamforming and ray tracing are performed in an urban environment. In this case, we use Dortmund downtown in Germany as the location and obtained its map from OpenStreetMap \cite{openstreetmap2026}, with the antenna placements shown in Fig. \ref{fig:antenna_placement}. The beamforming angle for each TX and RX is -60 to 60 degrees with a step size of 5 degrees, resulting in 25 values for each of the six antennas. Thus, the total number of combinations is $25^6=244,140,625$. Via exhaustive search, the maximum total channel capacity is $8.60$ Gbps.

\begin{table}[t]
	\centering
	\caption{TX and RX coordinates.}
	\vspace*{-0.3cm}
	\begin{tabular}{lccc}
		\hline
		Antennas & $x$ & $y$ & $z$ \\
		\hline
		TX1 & 58.1471 & -107.1382 & 8.1540 \\
		TX2 & -32.6813 & -36.6852 & 20.0856 \\
		RX1 & 114.6729 & -141.9017 & 1.8747 \\
		RX2 & -31.1958 & -68.3121 & 1.8394 \\
		RX3 & -26.5338 & 127.7987 & 1.6667 \\
		RX4 & 143.9960 & -89.3568 & 2.1643 \\
		\hline
	\end{tabular}\label{tab:table1}
	\vspace{-0.3cm}
\end{table}

TX and RX positions are modeled via a Poisson Point Process (PPP), with resulting locations shown in Table \ref{tab:table1}. To model the spatial distribution of users and base stations, we use a PPP to randomly generate TX and RX locations within the considered area. The Cartesian coordinates are then converted to geographic coordinates to enable map-based propagation and ray-tracing analysis. The channel is modeled using a MATLAB-based SBR method, considering up to one reflection and three diffractions, with concrete as the material.

The assignment of users to base stations is based on a game-theoretic approach considering channel conditions, noise, and shadowing. After assignment, TX and RX antenna patterns are modeled at different angles. The propagation channel is then computed using a ray-tracing model, where parameters such as path coefficients, phase, delay, and angles of arrival and departure are extracted. Antenna gains at both TX and RX sides are incorporated to compute the effective power of each path. The SINR for each user is calculated considering desired signal power, interference, and thermal noise. Finally, the total channel capacity is the sum of individual user capacities.

\section{Proposed Methodology}
This section is based on a hybrid framework consisting of three main components: RL, Bayesian inference, and game theory, which enables the system to function not merely as a numerical optimizer, but as a risk-aware, uncertainty-driven, and behavior-sensitive decision-maker. At a macro level, the learning process is designed as an iterative loop that observes the network state, generates an appropriate decision, evaluates its performance, and finally updates its knowledge. This structure follows the standard agent-environment interaction model in RL. An important aspect of this implementation is that the ``decision'' is not simply a choice among beamforming angles; rather, it is the outcome of interactions among several analytical layers, including a Bayesian model for risk estimation, a game-theoretic model for predicting the behavior of other RXs, and an RL model for long-term optimization. 

In this architecture, the network state is defined as a compact representation of the security and communication conditions of the system. This state includes the Bayesian-like belief that each user is malicious, as well as the system's confidence in these estimates. Thus, the system first attempts to transform the uncertain environment into a structured representation that can be utilized for learning. Since the quality of the state definition directly affects the quality of the final RL policy, this step is critically important. In practice, instead of directly observing the physical environment, the learning agent interacts with a probabilistic and layered representation of reality, which makes the decision-making process both more accurate and more robust to noise. This improvement is due to reduced sensitivity to noisy wireless measurements, explicit modeling of uncertainty in user classification, and incorporation of temporal information from multiple past observations. At the beginning of each iteration, the system attempts to identify the most suspicious node through a Bayesian probability vector, or belief state, which represents the probability of each user being an attacker such that, for $P_{j} = \Pr({\text{User}}_{j} = \text{Attacker})$, we have $\text{belief} = \left[ P_{1},P_{2},\ldots,P_{N} \right]$. Moreover, $\max \text{belief}$ represents the highest value in the belief vector and indicates the node with the greatest probability of being an attacker, reflecting the system’s confidence in detecting the suspicious node. The belief is updated using a Bayesian-like rule based on the deviation between predicted and observed capacity, with an exponential likelihood derived from the anomaly score. The resulting posterior is obtained through normalization and recursive smoothing over iterations.

However, the system's level of confidence in this diagnosis is also important. Thus, the obtained probabilities are converted into discrete confidence levels:
\begin{equation*}
    \text{confidence level} = \lfloor \max \text{belief} \times \text{numStatesPerNode}\rfloor + 1
\end{equation*}
such that the number of security levels is determined by the parameter numStatesPerNode, which is assumed to be 5, i.e., each node has 5 discrete levels of confidence, and the entire state space is transformed into a combination of node index (ID) and risk level. This discretization is applied, instead of using continuous probability, to reduce the complexity of the state space and stabilizing the Q-learning process, as converting continuous belief to finite levels both facilitates convergence and introduces the concept of ``security risk degree'' into the decision-making policy in a learnable way. The suspected node ID (suspectedNodeID) and confidence level are combined to construct the state
\begin{align}
&\text{State} = (\text{suspectedNodeID} - 1) \times \text{numStatesPerNode}\nonumber\\
&\qquad\qquad + \text{confidenceLevel}
\end{align}
given as the input to Q-learning. The action corresponds to selecting the beamforming angle for the TX and RX antennas, with action space of 25 predefined angles for each antenna. The RL agent's beam selection maximizes total channel capacity and minimizes interference and attacker effects.

To perform this selection, an $\epsilon$-greedy policy is used. However, unlike the classical RL framework, in this architecture, exploration and exploitation do not operate independently. Instead, these two phases are integrated with the Stackelberg-Bayesian game model. Thus, $\epsilon$-greedy only determines the level of randomness in the decision-making process, rather than directly generating the policy. Specifically, a uniformly distributed random variable $r \sim \mathcal{U}(0,1)$ is generated at each decision step to introduce randomness, such that if $r < \epsilon$, the system enters the exploration phase; otherwise, exploitation is performed. During the initial iterations, the beams are selected uniformly randomly to generate initial data for learning and to cover the state space. However, after the learning process begins, exploration is guided by the Stackelberg-Bayesian model. This model uses Bayesian information, channel state, antenna structure, and the predicted behavior of other actors to generate a meaningful search strategy. This feature places the architecture in the category of a hybrid decision policy.

When the system enters the exploitation phase, the action selection is based on maximizing the Q-value as $a^* = \arg\max_{a}Q(s,a)$ \cite{geranmayeh2024machine},
where $Q(s,a)$ represents the value of taking action $a$ in state $s$, and expresses the expected future reward obtained by selecting a particular beam. This decision is also aligned with the information obtained from the Stackelberg-Bayesian model. Thus, Q-learning plays the role of stabilizing the policy, while the game-theoretic model modifies the strategy. The Q-table update process is carried out according to the standard relationship \cite{geranmayeh2024machine}
\begin{align}
Q(s,a) \leftarrow Q(s,a) \!+\! \alpha\left[ r \!+\! \gamma\max_{a'}Q(s',a') \!-\! Q(s,a) \right],
\end{align}
where $\alpha$ is the learning rate, $\gamma$ is the discount factor for future rewards, $r$ is the received reward, and $s'$ is the next state of the system. This equation allows the RL agent to gradually learn which beams provide the highest total channel capacity and the lowest risk in the presence of attackers and varying channel conditions. To design system-level decision-making, two game-theoretic models were implemented and compared: the Nash game and the Stackelberg game. In both models, Bayesian information is directly incorporated into the utility function. The objective function is defined as
\begin{align}
U = \sum_{j}^{}{\left( 1 - P_{j} \right)C_{j} - \beta_{\text{risk}}P_{j}C_{j}},
\end{align}
where $C_j$ is the channel capacity of user $j{\in [1:N]}$, $P_j$ the probability that user $j$ is an attacker, and $\beta_\text{risk}$ the security penalty factor. This formulation ensures that both legitimate and potentially malicious users are jointly considered through a unified network-level metric. In the Nash-Bayesian model, each TX independently computes its best response to the current strategy of the other party. This process is performed using iterative best response and aims to reach a fixed point of the best-response iterations. The best-response condition is defined as $U_{1}\left( a_{1}^{*},a_{2}^{*} \right) \geq U_{1}\left( a_{1},a_{2}^{*} \right)$. 
The proposed attacker detection framework uses the aggregate network capacity as a network-level performance metric. This enables global system awareness while reducing computational overhead. Although based on per-user capacities, decision-making is performed at the network level, avoiding continuous per-user tracking and improving robustness to fading and shadowing. In this structure, both TXs act reactively and at the same hierarchical level. In environments with strong interference or active attackers, this method may get trapped in local equilibria and not be optimal in terms of total channel capacity. Thus, we also develop a Stackelberg-Bayesian model, where the decision-making structure is hierarchical, with one TX acting as the Leader and the other as the Follower. The Leader makes its decision first, and then the Follower computes its best response. The Leader role is also determined dynamically such that the TX serving more users is given decision-making priority. The Leader problem is defined as follows:
\begin{align}
&a_{L}^{*} = \arg\max_{a_L}U_{L}\left( a_{L},a_{F}^{*}\left( a_{L} \right) \right),\\
&a_{F}^{*}\left( a_{L} \right) = \arg\max_{a_F}U_{F}\left( a_{L},a_{F} \right).
\end{align}
The results obtained from the Stackelberg model are directly used in the utility function and the RL process. 

Next, we design two structures to update the Q-table, i.e., a memory-based model and a memoryless model, to investigate the effect of experience replay on the quality of learning. In the memory-based model, the agent saves its experiences in the form of quadruples as $( s,a,r,s')$. In the subsequent stages, these experiences are randomly replayed. This process reduces temporal correlations in the data, improves the stability of learning, and accelerates convergence. Furthermore, multiple updates are performed on the Q-table in each iteration, which enhances learning from important experiences (such as the occurrence of a severe attack). In contrast, in the memoryless model, the system uses only the current experience, and no replay mechanism is applied. This method is simpler but exhibits higher fluctuations and less stable performance in noisy and adversarial environments. At the end of each iteration, the exploration rate is gradually reduced to $\epsilon = \max(0.05,\epsilon \times 0.99)$. This process makes the system initially more exploratory and then gradually focused on the optimal learned policies, while the minimum value of $\epsilon$ ensures that the system never completely abandons exploration. Finally, this part can be considered the core of the system's decision-making process, where Bayesian information, game theory, and RL are integrated to produce the final beamforming decision in an intelligent and adaptive manner. Hence, the system can detect attackers and provide stable and optimal performance under complex and adversarial conditions, while suggesting a suitable angle for antenna beamforming.

\begin{figure}[t]
	\centering
	\includegraphics[width=0.9\columnwidth,keepaspectratio]{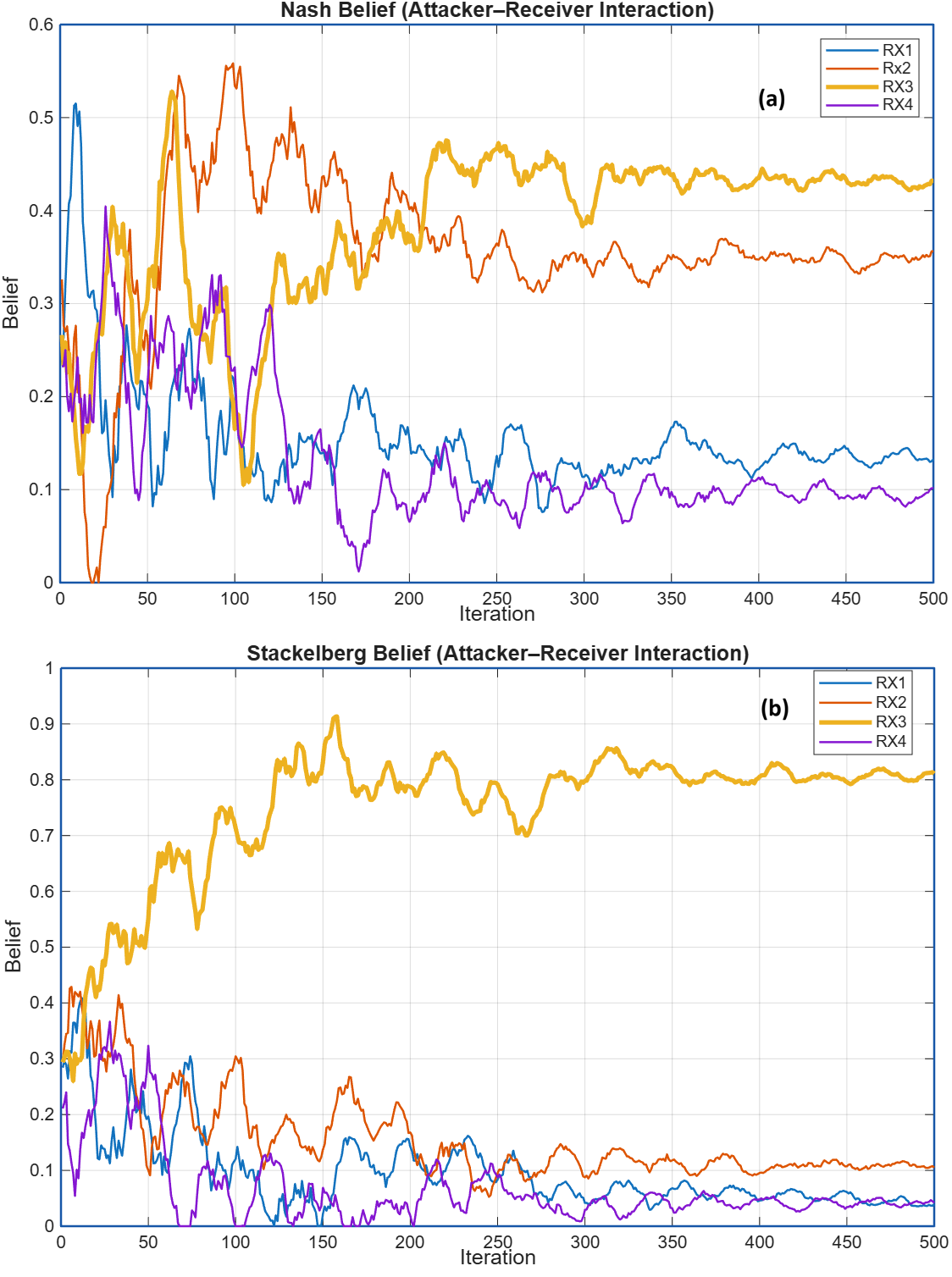}
	\vspace{-0.2cm}
	\caption{Evolution of Bayesian-like belief values for attacker detection under (a) Nash-type best-response optimization and (b) Stackelberg-type leader-follower optimization. The attacker is RX3.}
	\label{fig:bayesian_belief_values}
	\vspace{-0.4cm}
\end{figure}

\section{Performance Evaluation}
We compare the performance of a Nash-type best-response optimization and a Stackelberg-type leader–follower structure, both used within a game-theoretic framework, for attacker identification. Fig. \ref{fig:bayesian_belief_values}(a) illustrates the changes in the belief value over the iterations for the Nash model, while Fig. \ref{fig:bayesian_belief_values}(b) presents the same criterion for the Stackelberg model. At the beginning of the process, the belief value is initialized at 25\% for all RXs. The Bayesian algorithm then dynamically updates these values by observing variations in total channel capacity. In the Nash model, TXs make decisions simultaneously and independently, without any specific coordination. This increases internal network interference. Thus, the Bayesian algorithm faces difficulty in distinguishing whether network performance degradation is caused by an attacker or by interference generated by the TXs. As shown in Fig. \ref{fig:bayesian_belief_values}(a), the confidence values of different users remain close to each other, and even after 500 iterations, the system identifies the attacker with low confidence, whose value does not exceed 0.5. In contrast, in the Stackelberg model, the leader-follower structure among TXs reduces internal interference and creates a clearer decision-making environment for the Bayesian algorithm.

As depicted in Fig. \ref{fig:bayesian_belief_values}(b), the confidence value associated with the attacker RX3 increases quickly and exceeds 0.9. This demonstrates that the network successfully identifies the attacker with high confidence. Thus, in addition to increasing network capacity, the Stackelberg strategy improves the performance of the attacker detection system and accelerates the convergence process. In contrast, under the Nash-type best-response approach, uncontrolled interference introduces noise into the Bayesian estimation process and reduces the accuracy of attacker detection. Furthermore, according to the results average detection accuracy in the Stackelberg model is $69.41$\%, whereas in the Nash model it is $38.38$\%. Given the superior performance of the Stackelberg model, this model is employed in the next stage to detect attackers and determine the selected beamforming angle using machine learning methods.

\begin{figure}[t]
	\centering
	\includegraphics[width=0.9\columnwidth,keepaspectratio]{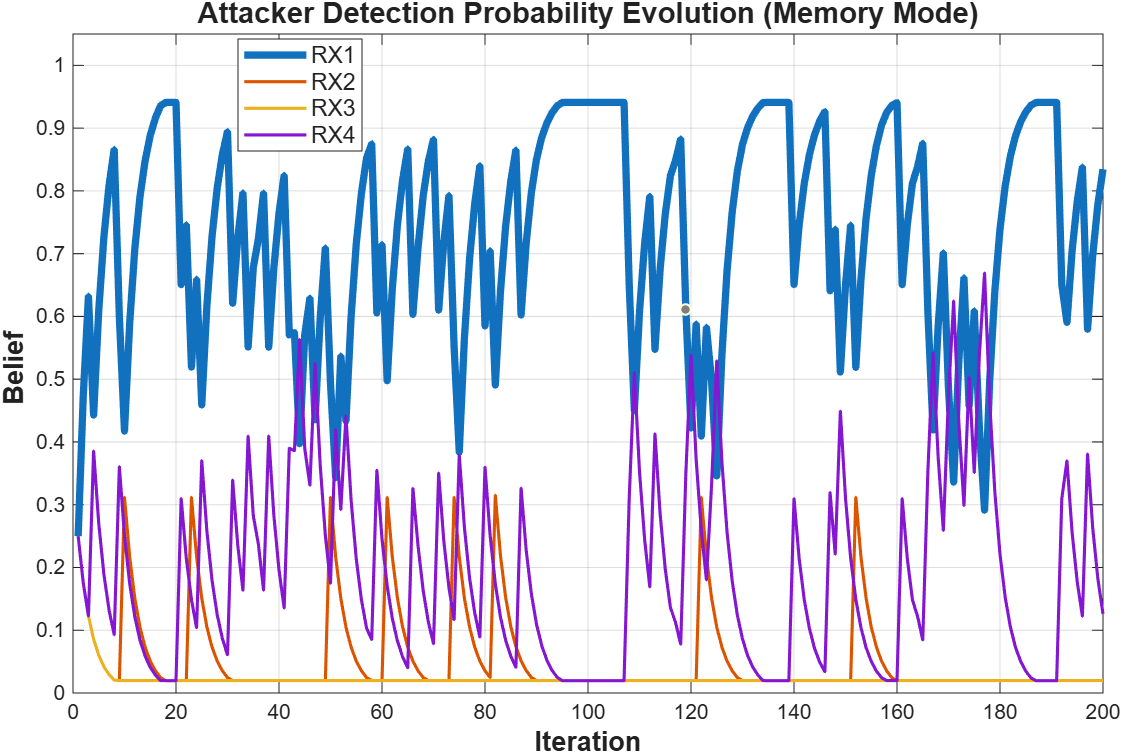}
	\caption{Attacker identification performance of Bayesian belief evolution in memory mode. }
	\label{fig:attacker_identification_memory}
	\vspace*{-0.4cm}
\end{figure}

In the next step, the Q-learning RL is employed to determine the optimal beamforming angle. The system performance is evaluated in two modes: ``With Memory'' and ``Without Memory''. These terms refer exclusively to the Q-learning mechanism and not to the state representation. The same state definition is used in both modes, but only the plots associated with the memory-based implementation are depicted. In the memory-based mode, the total channel capacity reaches an average of $934$ Mbps, whereas in the without memory mode it remains approximately $841$ Mbps, an improvement of about 11.06\%. This gain is attributed to the ability of network nodes to learn the attacker's behavior from past experiences using the Experience Replay mechanism, enabling faster adaptation compared to the memoryless case. The detection accuracy is $75.46$\% in the memory-based mode and $70.13$\% in the memoryless mode, indicating that memory improves the Bayesian estimator’s ability to distinguish malicious behavior from channel fading effects. However, this comes at a slightly higher computational cost, with execution times of $84$ seconds and $73$ seconds for the memory-based and memoryless modes, respectively, reflecting a trade-off between detection performance, delay, and throughput.
As the Stackelberg model combined with memory-based RL provides the best performance, its behavior is evaluated over 200 iterations. Next, the proposed framework is extended and the simulation is re-run using the updated architecture. The attacker is randomly selected in each run; in this case, RX1 is chosen as the attacker. Fig. \ref{fig:attacker_identification_memory} shows that the attacker RX1 is clearly separated from other users, indicating a reduced false alarm rate. Belief saturation is observed, where the belief for RX1 remains nearly constant at approximately 0.94 during iterations 20-30 and 100-110, confirming high-confidence detection. Temporary drops in belief occur due to changes in attacker behavior or channel conditions, but the system quickly recovers mainly due to memory, restoring stable detection. Occasional slight increases in the belief of other users are caused by strong interference and reflect realistic channel dynamics in 28 GHz mmWave environments with multipath and shadowing. Thus, the proposed Bayesian-RL framework achieves robust and stable detection, consistently identifying the attacker with probability above 0.9. Compared to the Nash-based approach, it provides faster recovery, lower false alarms, and improved stability.

\begin{figure}[t]
	\centering
	\includegraphics[width=0.9\columnwidth,keepaspectratio]{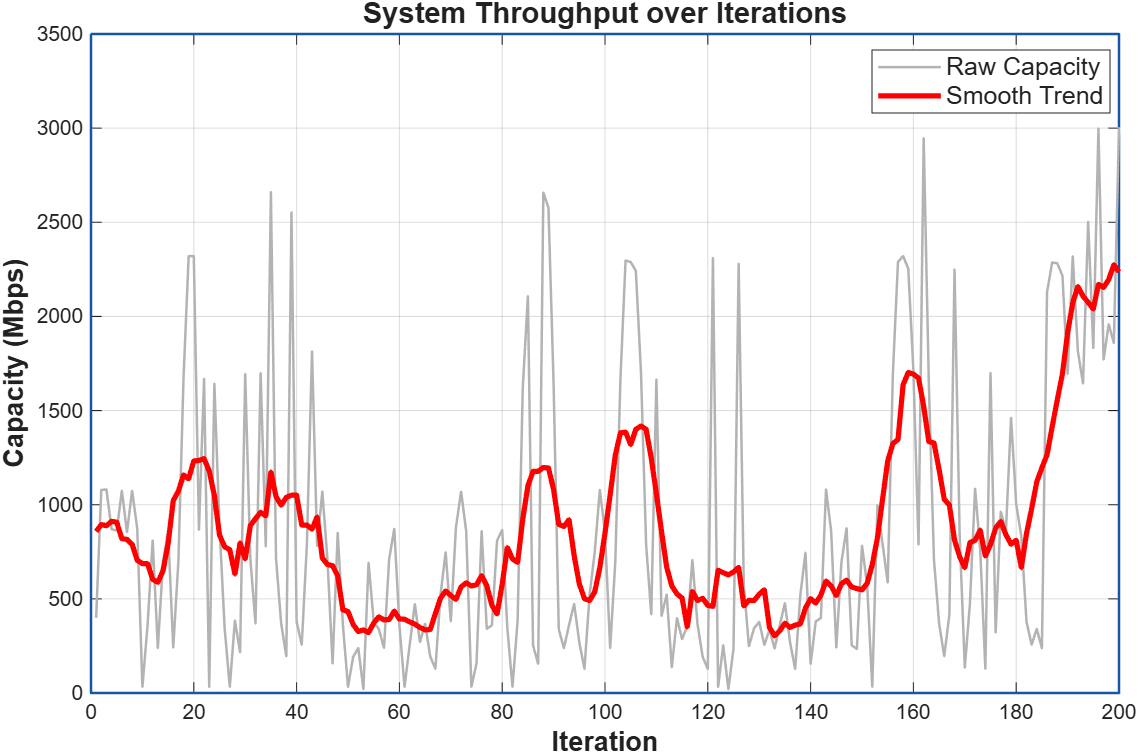}
	\caption{Evolution of system throughput in the presence of an attacker.}
	\label{fig:throughput_capacity_loss}
	\vspace*{-0.4cm}
\end{figure}

Fig. \ref{fig:throughput_capacity_loss} illustrates the system throughput and the total channel capacity degradation caused by an attacker that induces incorrect beamforming angles. In the initial iterations, total channel capacity fluctuates significantly due to uncertainty in attacker identification and severe shadowing effects in the mmWave environment. As the RL process progresses and Bayesian belief values converge, the system is able to select improved antenna configurations, allowing the total system capacity to reach approximately $2500$ Mbps despite the presence of an attacker. This value is consistent with the requirements of 5G New Radio systems and indicates efficient use of communication resources. The increasing trend in the final phase demonstrates policy improvement in the reported simulation and overall improvement in system performance. Consequently, the quality of service (QoS) for legitimate users is improved. This result confirms the effectiveness of the combined Stackelberg-Bayesian approach for radio resource management in adversarial environments.

\section{Conclusion}
We investigated the use of RL, Bayesian inference, and game theory for attacker detection and beamforming optimization in dynamic 6G ISAC systems. A Bayesian game-theoretic framework was developed to model uncertainty in TX-RX interactions, and the Stackelberg model was shown to provide more stable and effective attacker detection than the Nash model. The Stackelberg outputs were integrated into a Q-learning process, where the memory-based update mechanism achieved higher stability, throughput, and detection accuracy than the memoryless one. Thus, the proposed framework improved attacker-aware beamforming performance and provided a basis for extensions to mobile and real-time ISAC scenarios.

\section*{Acknowledgment}
This work was supported by the German Federal Ministry of Research, Technology and Space (BMFTR) 6GEM+ Transfer Hub under Grants 16KIS2412 and 16KISS005.

\bibliographystyle{IEEEtran}
\bibliography{bibliofile}

\begin{thebibliography}{10}
\providecommand{\url}[1]{#1}
\csname url@samestyle\endcsname
\providecommand{\newblock}{\relax}
\providecommand{\bibinfo}[2]{#2}
\providecommand{\BIBentrySTDinterwordspacing}{\spaceskip=0pt\relax}
\providecommand{\BIBentryALTinterwordstretchfactor}{4}
\providecommand{\BIBentryALTinterwordspacing}{\spaceskip=\fontdimen2\font plus
\BIBentryALTinterwordstretchfactor\fontdimen3\font minus
  \fontdimen4\font\relax}
\providecommand{\BIBforeignlanguage}[2]{{%
\expandafter\ifx\csname l@#1\endcsname\relax
\typeout{** WARNING: IEEEtran.bst: No hyphenation pattern has been}%
\typeout{** loaded for the language `#1'. Using the pattern for}%
\typeout{** the default language instead.}%
\else
\language=\csname l@#1\endcsname
\fi
#2}}
\providecommand{\BIBdecl}{\relax}
\BIBdecl

\bibitem{mamaghani2026securing}
M.~T. Mamaghani, X.~Zhou, N.~Yang, and A.~Lee~Swindlehurst, ``Securing
  integrated sensing and communication against a mobile adversary: {A
  Stackelberg} game with deep reinforcement learning,'' \emph{IEEE Journal on
  Selected Areas in Communications}, vol.~44, pp. 942--958, 2026.

\bibitem{mukherjee2014principles}
A.~Mukherjee, S.~A.~A. Fakoorian, J.~Huang, and A.~L. Swindlehurst,
  ``Principles of physical layer security in multiuser wireless networks: {A}
  survey,'' \emph{IEEE Communications Surveys \& Tutorials}, vol.~16, no.~3,
  pp. 1550--1573, 2014.

\bibitem{OnurRoleofISAC}
M.~U.~F. Qaisar \emph{et~al.}, ``The role of {ISAC in 6G Networks: Enabling}
  next-generation wireless systems,'' \emph{IEEE Transactions on Network
  Science and Engineering}, vol.~13, pp. 7825--7861, 2026.

\bibitem{zou2024energy}
J.~Zou \emph{et~al.}, ``Energy-efficient beamforming design for integrated
  sensing and communications systems,'' \emph{IEEE Transactions on
  Communications}, vol.~72, no.~6, pp. 3766--3782, 2024.

\bibitem{he2023full}
Z.~He \emph{et~al.}, ``Full-duplex communication for {ISAC: Joint} beamforming
  and power optimization,'' \emph{IEEE Journal on Selected Areas in
  Communications}, vol.~41, no.~9, pp. 2920--2936, 2023.

\bibitem{zhang2025research}
Y.~Zhang \emph{et~al.}, ``Research and experimental validation for {3GPP ISAC}
  channel modeling standardization,'' \emph{{Elsevier} Digital Communications
  and Networks}, 2025.

\bibitem{OnurSecureISACJournal}
O.~Günlü, M.~R. Bloch, R.~F. Schaefer, and A.~Yener, ``Secure integrated
  sensing and communication,'' \emph{IEEE Journal on Selected Areas in
  Information Theory}, vol.~4, pp. 40--53, 2023.

\bibitem{Onur2026secureISACTutorial}
T.~Welling, O.~G{\"u}nl{\"u}, and A.~Yener, ``Secure integrated sensing and
  communication: {Information} theory offers insights,'' \emph{arXiv preprint
  arXiv:2605.08106}, 2026.

\bibitem{maheshwari2019flexible}
M.~K. Maheshwari, M.~Agiwal, N.~Saxena, and A.~Roy, ``Flexible beamforming in
  {5G} wireless for internet of things,'' \emph{IETE technical review},
  vol.~36, no.~1, pp. 3--16, 2019.

\bibitem{geranmayeh2024comparison}
P.~Geranmayeh and E.~Grass, ``Comparison of optimization techniques and machine
  learning methods for optimized beamforming in wireless networks,'' in
  \emph{{IEEE-APS} Topical Conference on Antennas and Propagation in Wireless
  Communications}, 2024, pp. 66--70.

\bibitem{Onur2026ISACfeedback}
Y.~Zhou, N.~Devroye, and O.~G{\"u}nl{\"u}, ``Feedback lunch: {Learned} feedback
  codes for secure communications,'' in \emph{ACM Workshop on Wireless Security
  and Machine Learning}, 2026.

\bibitem{lei2018incomplete}
C.~Lei, H.-Q. Zhang, L.-M. Wan, L.~Liu, and D.-h. Ma, ``Incomplete information
  {Markov} game theoretic approach to strategy generation for moving target
  defense,'' \emph{{Elsevier} Computer Communications}, vol. 116, pp. 184--199,
  2018.

\bibitem{jiang2021markov}
L.~Jiang, H.~Zhang, and J.~Wang, ``Optimal decision-making method for moving
  target defense based on multi-stage {Markov} signal game,'' \emph{Acta
  Electron. Sin}, vol.~49, pp. 527--535, 2021.

\bibitem{teofilo2012adapting}
L.~F. Te{\'o}filo, N.~Passos, L.~P. Reis, and H.~L. Cardoso, ``Adapting
  strategies to opponent models in incomplete information games: {A}
  reinforcement learning approach for poker,'' in \emph{{Springer}
  International Conference on Autonomous and Intelligent Systems}, 2012, pp.
  220--227.

\bibitem{ren2023cooperative}
Z.~Ren, D.~Zhang, S.~Tang, W.~Xiong, and S.-h. Yang, ``Cooperative maneuver
  decision making for multi-{UAV} air combat based on incomplete information
  dynamic game,'' \emph{{Elsevier} Defence Technology}, vol.~27, pp. 308--317,
  2023.

\bibitem{yao2024bayesian}
P.~Yao, Z.~Jiang, B.~Yan, Q.~Yang, and W.~Wang, ``Bayesian and stochastic game
  joint approach for cross-layer optimal defensive decision-making in
  industrial cyber-physical systems,'' \emph{{Elsevier} Information Sciences},
  vol. 662, p. 120216, 2024.

\bibitem{ETSI_TR_138901_v1610}
{ETSI}, ``{5G; Study on channel model for frequencies from 0.5 to 100 GHz (3GPP
  TR 38.901 version 16.1.0 Release 16)},'' Tech. Rep., Nov. 2020.

\bibitem{mathworks2026phased}
{MathWorks}, \emph{Phased Array System Toolbox User's Guide}, Natick, MA, USA,
  2026.

\bibitem{itur2009m2135}
{ITU-R}, ``Guidelines for evaluation of radio interface technologies for
  {IMT-Advanced},'' Geneva, Switzerland, Tech. Rep. M.2135-1, Dec. 2009.

\bibitem{openstreetmap2026}
{OpenStreetMap contributors}, ``Openstreetmap,'' n.d., accessed: April 2026.

\bibitem{geranmayeh2024machine}
P.~Geranmayeh and E.~Grass, ``Machine learning based beam selection for
  maximizing wireless network capacity,'' \emph{IEEE Access}, vol.~12, pp.
  45\,176--45\,186, 2024.

\end{thebibliography}

\end{document}